\documentclass[10pt,preprint2,deluxetable]{aastex}

\usepackage{color}
\definecolor{citeRGB}{rgb}{0,0.1,0.7}
\usepackage[hyperfootnotes=true,naturalnames=true,letterpaper,pdfstartview=FitH,pdfpagemode=UseNone,colorlinks=true,citecolor=citeRGB]{hyperref}

\gdef\survey{GO-14114 }
\gdef\name{COSMOS-DASH }
\shortauthors{Momcheva et al.}
\shorttitle{Wide-field near-IR imaging with HST}
\slugcomment{Submitted to Publications of the Astronomical Society of the Pacific}
\begin{document}

\newcommand\XXX[1]{{\textcolor{red}{\textbf{x\ #1\ x}}}}
\newcommand\XX{{\textcolor{red}{\textbf{XX}}}}

\title{A New Method for Wide-Field Near-IR Imaging with the Hubble Space Telescope}

\author{Ivelina G. Momcheva,\footnote{Space Telescope Science Institute, Baltimore, MD 21218, USA}\,$^{,}$\footnote{\href{mailto:imomcheva@stsci.edu}{imomcheva@stsci.edu}} \,
Pieter G.\ van Dokkum,\footnote{Department of Astronomy, Yale University, 260 Whitney Avenue, New Haven, CT, USA 06511} \,
Arjen van der Wel,\footnote{Max Planck Institute for Astronomy, D-69117, Heidelberg, Germany} \,
Gabriel B. Brammer,\altaffilmark{1} \,
John Mackenty,\altaffilmark{1}\,
 Erica J. Nelson,\altaffilmark{3}\,
 Joel Leja,\altaffilmark{3} \,
 Adam Muzzin,\footnote{Institute of Astronomy, University of Cambridge, Cambridge, UK} \,\,
 Marijn Franx\footnote{Leiden Observatory, Leiden University, Leiden, The Netherlands}
 }

%\altaffiltext{1}{Space Telescope Science Institute, Baltimore, MD 21218, USA}
%\altaffiltext{2}{Department of Astronomy, Yale University, 260 Whitney Avenue, New Haven, CT, USA 06511}
%\altaffiltext{3}{Max Planck Institute for Astronomy, D-69117, Heidelberg, Germany}
%\altaffiltext{4}{Institute of Astronomy, University of Cambridge, Cambridge, UK}
%\altaffiltext{5}{Leiden Observatory, Leiden University, Leiden, The Netherlands}

\begin{abstract}
We present a new technique for wide and shallow observations using the near-infrared channel of Wide Field Camera 3 (WFC3) on the {\em Hubble Space Telescope} (HST). Wide-field near-IR surveys with HST are
generally inefficient, as guide star acquisitions make it impractical
to observe more than one pointing per orbit. This  limitation
can  be  circumvented  by  guiding  with  gyros  alone,  which  is  possible  as  long  as  the  telescope
has  three  functional  gyros. The method presented here allows us to observe mosaics of eight independent WFC3-IR
pointings in a single orbit by utilizing the fact that HST
drifts by only a very small amount in the 25 seconds between
non-destructive reads of unguided exposures. By shifting the reads and
treating them as independent exposures the full resolution of WFC3 can
be restored.  We use this ``drift and shift'' (DASH) method in the Cycle
23 \name program, which will obtain 456
WFC3 $H_{160}$ pointings in 57 orbits, covering an area of 0.6
degree$^2$ in the COSMOS field down to $H_{160} = 25$. When completed,
the
program will more than triple
the area of extra-galactic survey fields covered by
near-IR imaging at HST resolution. We demonstrate
the viability of the method with the
first four orbits (32 pointings) of this program. We show that the
resolution of the WFC3 camera is preserved, and that structural
parameters of galaxies are consistent with those measured in guided
observations.

\end{abstract}


\section{Introduction}
Over its lifetime the
{\em Hubble Space Telescope} (HST) has imaged many ``blank'' fields
at many wavelengths, to obtain a census of the Universe over most of
its history.
The survey strategy of the extra-galactic community has
been to image a few individual HST pointings to great depth \citep[examples
are the
Hubble Deep Fields, the Ultra Deep Field, and the Frontier Fields; ][]{hdf,ellis:13,illingworth:13,lotz:14},
and larger areas to shallower depth \cite[the GEMS
survey, COSMOS, the GOODS North and South fields, CANDELS, etc.; ][]{rix:04,scoville:07,goods,grogin:11,koekemoer:11}.
This ``wedding cake'' strategy of tiered surveys
is driven by the form of the luminosity function of most astronomical
objects. The number density of faint objects is almost always
larger than that of
bright objects, which means that representative samples of faint
objects can be obtained in deep, pencil-beam surveys and representative
samples of bright objects in shallow, wide-area surveys.

Ground-based surveys have extended the shallow, wide-area domain to
degree-scales and beyond: 
SDSS covers $\sim 1/3$ of the entire sky in the optical,
and a plethora of optical and near-IR surveys are covering areas up to
thousands of square degrees \citep*[e.g., the 155 degree$^2$ CFHTLenS survey, the
ESO Kilo Degree Survey, and the 5000 degree$^2$ Dark Energy Survey
in the optical, and the tiered UKIDSS near-IR surveys: the 12
degree$^2$ VIDEO survey, and the deep 1 degree$^2$ UltraVISTA survey in the
near-IR; ][]{cfhtlens,kids:07, kids:13,des,video,ultravista}. These surveys measure the high mass end of the galaxy
mass function at $0<z<4$ and also
address questions such as the number density of
bright Lyman break galaxies out to $z\sim 8$, the clustering
of galaxies, the evolution of galaxy groups and clusters,
the properties and demographics of AGNs, and the prevalence
of short-lived events such as mergers.

Many of these science questions would benefit greatly
from imaging at HST resolution.
However, the widest/shallowest tiers of the wedding cake are
devoid of HST imaging, particularly in the near-IR.
The largest area imaged with HST in the optical is
the 2 degree$^2$ COSMOS field, which was
carried out with ACS in the $I_{814}$
filter in Cycles 12 and 13
\citep{scoville:07} at a cost of 640 orbits.
By contrast,
the widest area imaged in the near-IR
is an order of magnitude smaller. The five fields of the
900-orbit CANDELS survey,  which used the $J_{125}$ and $H_{160}$
filters of the WFC3 camera in a Multi-Cycle Treasury program
in Cycles 18, 19, and 20, add up to
0.25 degree$^2$.

The reason for the lack of very wide HST near-IR
surveys seems obvious:
the price of HST's excellent resolution and small pixels
is a small field of view, and a single WFC3 pointing
covers a mere 4.6 arcmin$^2$. However, this explanation is not sufficient,
as ground based surveys routinely cover thousands of times their detector
area. As an example, the UKIDSS Large Area Survey (LAS)
covers 4000 degree$^2$ in $Y$, $J$, $H$, and $K$
using $\sim 20,000$ pointings with
the 0.21 degree$^2$ WFCAM instrument. An HST/WFC3 survey with a similar
strategy (that is, a similar number of pointings) would
cover 25 degree$^2$.  Furthermore, as the
sensitivity of WFC3 is similar to that of a 30\,m telescope on
the ground, even short exposures reach depths comparable to the deepest ground-based surveys in existence: a 300\,s
exposure in $H_{160}$ gives a $5\sigma$
point source sensitivity of AB\,=\,25.4, comparable to
the deepest ground-based surveys in existence.

The real limitation when designing wide-field programs
is that HST has a natural lower limit to the exposure time per pointing,
which stems from
the time that is required for a guide star acquisition and other
overheads associated with moving the telescope.
The UKIDSS Large Area Survey
has an exposure time of 40\,s per band, and this would be hopelessly
inefficient with HST: even if many guide star acquisitions were allowed
in a single orbit
(the limit is two), only four would fit and nearly the entire
$\sim 50$\,minutes of orbital visibility would be taken up with overhead.
As a result, the natural lower limit to the exposure time
is a single orbit,\footnote{Note that it is possible
to split the orbit into shorter observations
with different filters; this reduces the per-filter observing time
but does not increase the area of the survey.}
and the natural upper limit to the area of an HST
survey is the number of orbits of the program multiplied by
the detector area of the instrument. 

\section{``Drift and Shift'': A New Method for Wide-Field WFC3 Surveys}

\subsection{Overview}
There is a way to circumvent the limitations imposed by the guide
star acquisitions. If no new guide star is acquired between pointings
the overheads decrease dramatically, and as we show below it is possible
to fit eight distinct pointings, each with an exposure time of
$\sim 300$\,s, in a single orbit.

\begin{figure*}[htbp]
\epsscale{2.}
\plotone{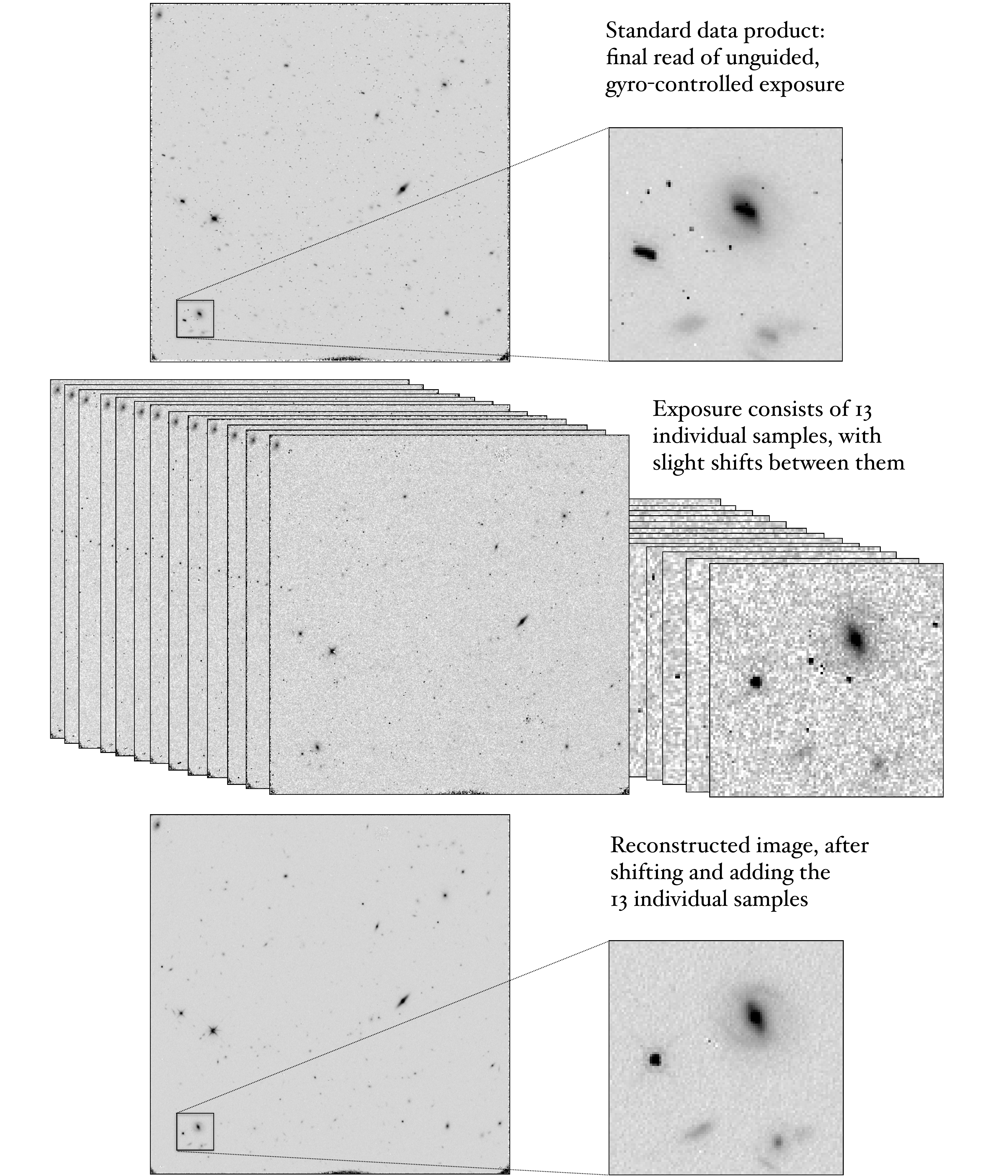}
\caption{\small
Illustration of the ``drift and shift'' (DASH) method of restoring unguided
HST images. The top panel and inset show the standard data product
(the \textsc{FLT} file) of an unguided, gyro-controlled
exposure. The objects are smeared due to the lack of fine guidance
sensor corrections. The middle panels show the twelve individual
samples that comprise the exposure, created from the non-destructive
reads. The smearing is small in each individual sample. The bottom
panels show the reconstructed image after shifting the twelve
samples to a common frame and adding them.}
\label{method.fig}
\end{figure*}

Operating without a guide star does not
change the telescope control: ever
since the last servicing mission the pointing of HST has
been controlled by three gyros, and this is the case irrespective
of whether a guide star is acquired or not. In a
standard guided exposure the three
gyros receive continuous corrections from the Fine Guidance Sensors
(FGS), and turning off guiding merely stops the stream of corrections
from the FGS.
The effect of the lack of corrections is that the telescope begins to drift
by an expected $0\farcs 001$ -- $0\farcs 002$
per second. The
drift can be larger during orbits when the telescope experiences unusually strong
atmospheric drag.\footnote{This is an expectation; there
are no systematic measurements of this effect because all 
observations (with the exception of spatial scans for bright objects)
are done under  FGS control.}

The result of operating in a gyro-only mode is therefore that exposures
longer than a minute
are smeared.  For CCD detectors, such images are scientifically unusable.  When a guide star
acquisition fails, or a guide star is lost during an observation,
the visit is flagged and typically such observations 
are repeated. However, a property of the WFC3
detector (and IR detectors in general) is that an exposure is composed
of multiple non-destructive, zero-overhead reads.  The exposure time
of an individual read can be set,\footnote{To certain specified values;
e.g., 10\,s, 25\,s, 50\,s, 100\,s, and 200\,s.}
and for times up to 25\,s the drift
{\em in between reads} is $\lesssim 0\farcs 05$, or less than half a
$0\farcs 13$ pixel.
The dataset obtained in the interval between two
reads is simply the difference between read $i$ and read $i-1$.
Therefore,  an unguided,
gyro-controlled, 300\,s exposure with 25\,s reads effectively consists of 12
independent, dithered exposures that can be shifted and combined into
a full resolution image with hot pixels and cosmic rays removed.
A schematic of this procedure is shown in Figure \ref{method.fig}.

After the 300\,s exposure the telescope can then be offset to
a new position without the need to acquire a guide star at that position;
as soon as the telescope move is completed the next integration can begin.
As we detail below, we can observe eight distinct positions in
a single orbit with this method.

\begin{figure*}[htbp]
\centering
\begin{minipage}[b]{0.75\linewidth}{
\includegraphics[width=\textwidth]{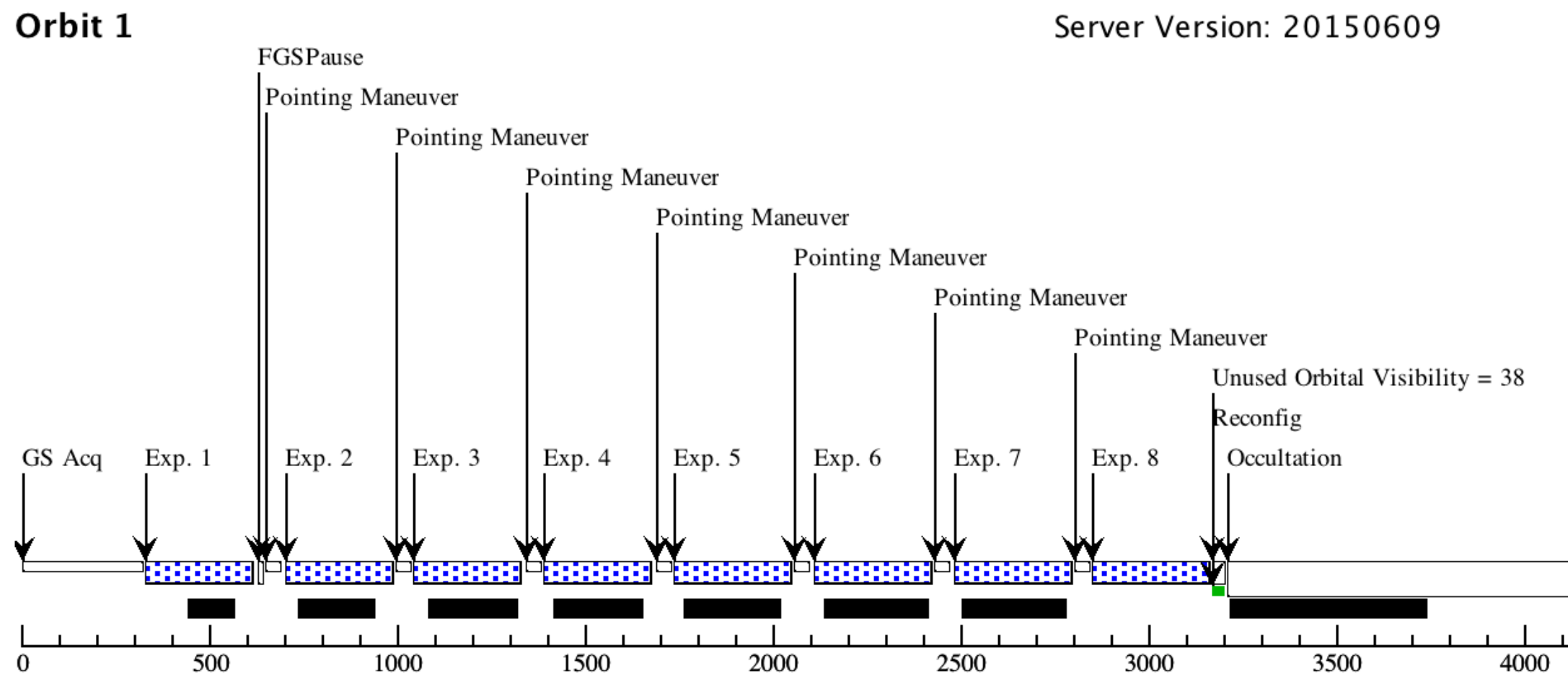}}
\end{minipage}
\begin{minipage}[b]{0.2\linewidth}{
\includegraphics[width=\textwidth]{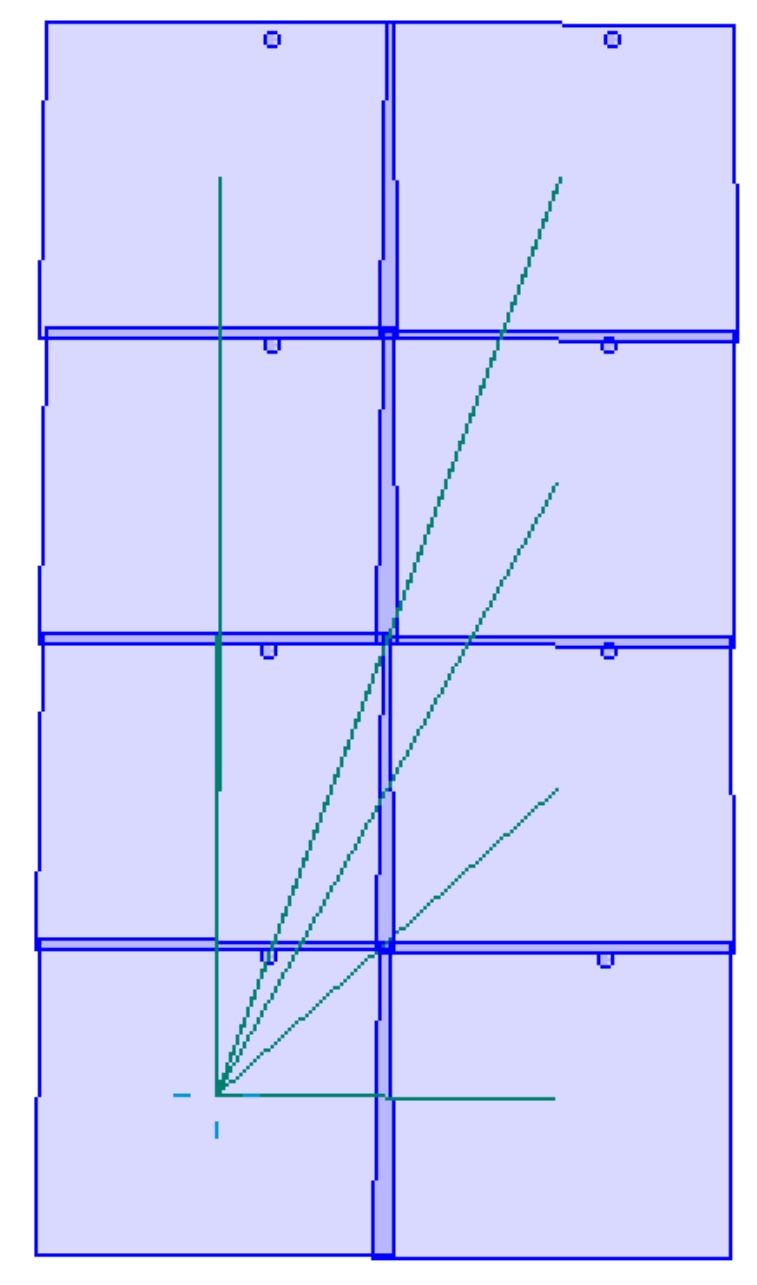}}
\end{minipage}
\caption{\small
Left: graphical representation of a single orbit (Visit 1
of GO-14144). Only the first pointing is
guided ({\tt PCS Mode=FINE}). Blue, dotted bars indicate science
exposures. Black bars are buffer dumps. The total exposure time is 2124\,s,
corresponding to 66\,\% of the total orbital visibility. Right: pattern
of the eight WFC3 pointings observed in this orbit. The green lines illustrate the shifts relative to the first pointing in the orbit. When the observations are carried out we step from one pointing to the next, without returning to the starting position.}
\label{orbit.fig}
\end{figure*}

\begin{deluxetable}{cllcc}
\label{tab:struct}
\tablecaption{Structure of a Single GO-14144 Orbit}
\tabletypesize{\footnotesize}
\tablewidth{0pt}
\tablehead{\colhead{Step} & \colhead{Event} & \colhead{Changes to keywords} &
\colhead{Duration} & \colhead{Exposure time}}
\startdata
1 & guide star acquisition  & \nodata & 333\,s & \\
2 & guided exposure, position 1 & {\tt PCS Mode=FINE} & 295\,s & 253\,s \\
   & & {\tt SAMP-SEQ=SPARS25} & & \\
   & & {\tt NSAMP=11} & & \\
3 & stop FGS corrections & \nodata  & 21\,s &\\
4 & offset to position 2 & \nodata  & 52\,s & \\
5 & unguided exposure, position 2 & {\tt PCS Mode=GYRO} & 295\,s & 253\,s \\
6 & offset to position 3 & \nodata  & 50\,s & \\
7 & unguided exposure, position 3 & \nodata   & 295\,s & 253\,s \\
8 & offset to position 4 &  \nodata & 52\,s & \\
9 & unguided exposure, position 4 &   \nodata & 295\,s & 253\,s \\
10 & offset to position 5 &  \nodata & 50\,s & \\
11 & unguided exposure, position 5 & {\tt NSAMP=12} & 320\,s & 278\,s \\
12 & offset to position 6 &  \nodata & 52\,s & \\
13 & unguided exposure, position 6 &  \nodata  & 320\,s & 278\,s \\
14 & offset to position 7 &  \nodata & 50\,s & \\
15 & unguided exposure, position 7 &   \nodata & 320\,s & 278\,s \\
16 & offset to position 8 & \nodata  & 52\,s & \\
17 & unguided exposure, position 8 & \nodata   & 320\,s & 278\,s \\
\hline
& Unused orbital visibility:  & & 38\,s & \\
& Total duration and exposure time: &  & 3209\,s & 2124\,s 
\enddata
\end{deluxetable}

\subsection{Implementation: Structure of a Single Orbit}
The optimal way to implement the ``drift and shift'' strategy is
dictated by the amount of data that can be stored in memory during
the exposure. We
first consider the minimum time between reads during
an exposure.\footnote{We used the Astronomer's Proposal Tool Phase
II software (\url{http://www.stsci.edu/hst/proposing/apt}) for this
analysis, and implemented this strategy in our GO-14144 program
described in \S \ref{survey.sec}.}
With 10\,s intervals between reads (SPARS10 mode) the buffer fills
so quickly that
memory dumps have to be conducted during the exposure, drastically reducing
the observing efficiency and negating the benefits of the method.
Fortunately, 25\,s
intervals (SPARS25 mode) are possible, and the typical drift
in that time is still significantly smaller than a pixel. For our purposes, longer
intervals
do not provide significant benefits in terms of the total on-target
exposure time, reduce the number of independent samples that are
available for cosmic ray rejection, and begin to show appreciable drift
within each sample. However, for other scientific applications such longer
intervals
may be more advantageous. 

Next we determine how many independent pointings we can fit in a single
orbit, by varying the number of samples taken during a single exposure (i.e.,
the per-pointing exposure time) and the number of pointings. This
calculation depends on the field that is observed, as that determines
the length of the visibility window. We only consider the COSMOS field
(at RA=$2^h15^m00^s$ and Dec=$10^{\circ}00'00"$), which is the target area of our GO-14144
program (see \S\ref{survey.sec}) and is typical for most of the
sky.\footnote{It is possible to fit more pointings in
continuous viewing zone observations, at the expense of an increased
background in some of the pointings and possibly larger drift rates.}
The maximum number of pointings is eight: when a larger number
is attempted (with a smaller number of samples taken during each exposure),
buffer dumps cause the ninth pointing to spill over into a second orbit.
The structure of the orbit is summarized in Table \ref{tab:struct} and shown
graphically in the left panel of
Figure \ref{orbit.fig}.  The pointing pattern
for this particular orbit (Visit 1 of GO-14144) is shown at right
in Figure \ref{orbit.fig}: the total covered area is $\sim 8\times$
greater than that of a single WFC3 pointing.

The exposure time per pointing is approximately
the exposure time per sample multiplied
by the number of samples. The number of samples is 11 ({\tt NSAMP=11})
for the first four pointings and 12 ({\tt NSAMP=12}) for the last four, however the first sample is read out  at the very beginning of the exposure (at 2.9 seconds).
The realized per-pointing exposure times are 253\,s and 278\,s
respectively. The total number of independent 25-second observations
obtained in this single orbit is $4\times 10 + 4\times 11 = 84$. The observing efficiency, expressed as the on-target
realized exposure time divided by the total orbital visibility, is
66\,\%, similar to the typical efficiency of standard
observing modes.  

\begin{figure*}[htbp]
\epsscale{1.9}
\plotone{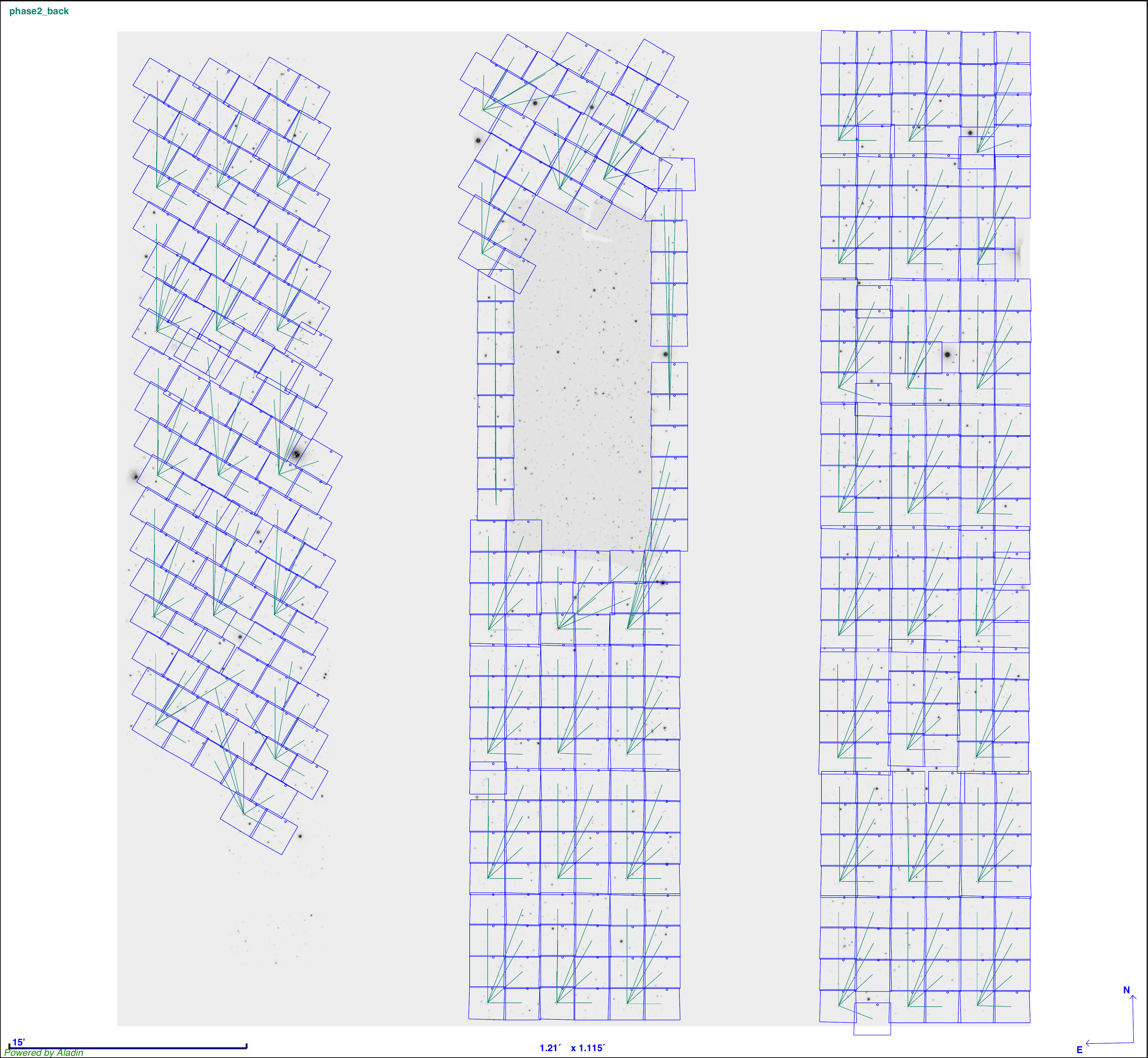}
\caption{\small
Layout of the 456 pointings of the \name program. The
background image is a composite of the UltraVISTA deep $H$ band imaging
and the CANDELS $H_{160}$ WFC3 imaging \citep[the ``hole'' in the
mosaic above the center of the middle
stripe;][]{grogin:11,koekemoer:11}. Small gaps between pointings are deliberate: they coincide
with bright sources that would cause persistence problems.}
\label{mosaic.fig}
\end{figure*}

\section{A Wide-Field Survey of the COSMOS Field -- \name}
\label{survey.sec}
We have undertaken a survey with the drift-and-shift technique during
Cycle 23, adding a wide-field tier to the ``wedding cake'' of near-IR
imaging surveys with HST. Covering 0.6 degree$^2$, \name (for ``Drift And SHift") more
than triples the extragalactic survey area that HST has observed in the
near-IR.  Here we describe the observing strategy
of this program (GO-14114); in \S\ \ref{demo.sec}
we analyze the first four visits
and demonstrate that the drift-and-shift technique
produces the expected results.

The 57-orbit \name program is targeting the COSMOS field \citep{scoville:07},
as this is the only field with optical ACS imaging (in the
$I_{814}$ filter) over a
sufficiently large contiguous area. The longest wavelength filter,
$H_{160}$, is used to maximize the color baseline at HST resolution.
We do not cover the entire 2 degree$^2$ that has ACS imaging
but only the
UltraVISTA deep stripes \citep{ultravista}; these
regions have extremely deep complementary ground-based $Y$, $J$, $H$,
and $K$ imaging as well as deep Spitzer IRAC imaging from the SMUVS
Exploration Science program (Spitzer GO-11016, PI: K. Caputi).

The layout of the $57\times 8 = 456$ pointings is shown in Figure
\ref{mosaic.fig}. The data are taken at two orientations to facilitate
scheduling. The large gap in the mosaic in the central stripe is the
area of the COSMOS field that was observed in $J_{125}$ and
$H_{160}$ in the
CANDELS Multi-Cycle program \citep{grogin:11,koekemoer:11}. We ensured that some pointings
partially overlap with CANDELS in order to test our ability to recover
the photometry and structural parameters of galaxies (see \S\ \ref{demo.sec}).
Small gaps in the mosaic are deliberate, and coincide with very bright
stars or galaxies. Those objects would cause severe persistence, with
the potential to affect the other pointings in an orbit as well as consecutive orbits.
The total area covered by \survey\ is 0.59 degree$^2$, which can be
compared to the existing total five-field CANDELS area of 0.24 degree$^2$.
\survey\ therefore increases the survey area that has WFC3 imaging in addition
to deep ground- and Spitzer data by a factor of 3.5.

The depth that will be achieved depends on the realized drift rate.
If the drift during a pointing is $\lesssim 1$ pixel the limiting
magnitudes will be identical to those of regular $253$\,s and
278\,s exposures. With the average zodaical background in the
COSMOS field the $5\sigma$ point source limit will then be $H_{160}=25.1$
(on the AB system).
If the drift rate is several pixels the point source
depth will
be slightly lower, as the shifts place different independent pixels onto
the same output pixel; a conservative expectation is $H_{160} \approx 24.9$
in those circumstances.
As we show in the next Section the drifts in the first four orbits
of our program are of order 1 pixel per pointing.

\begin{figure}[htbp]
\centering
\includegraphics[width = 0.45\textwidth]{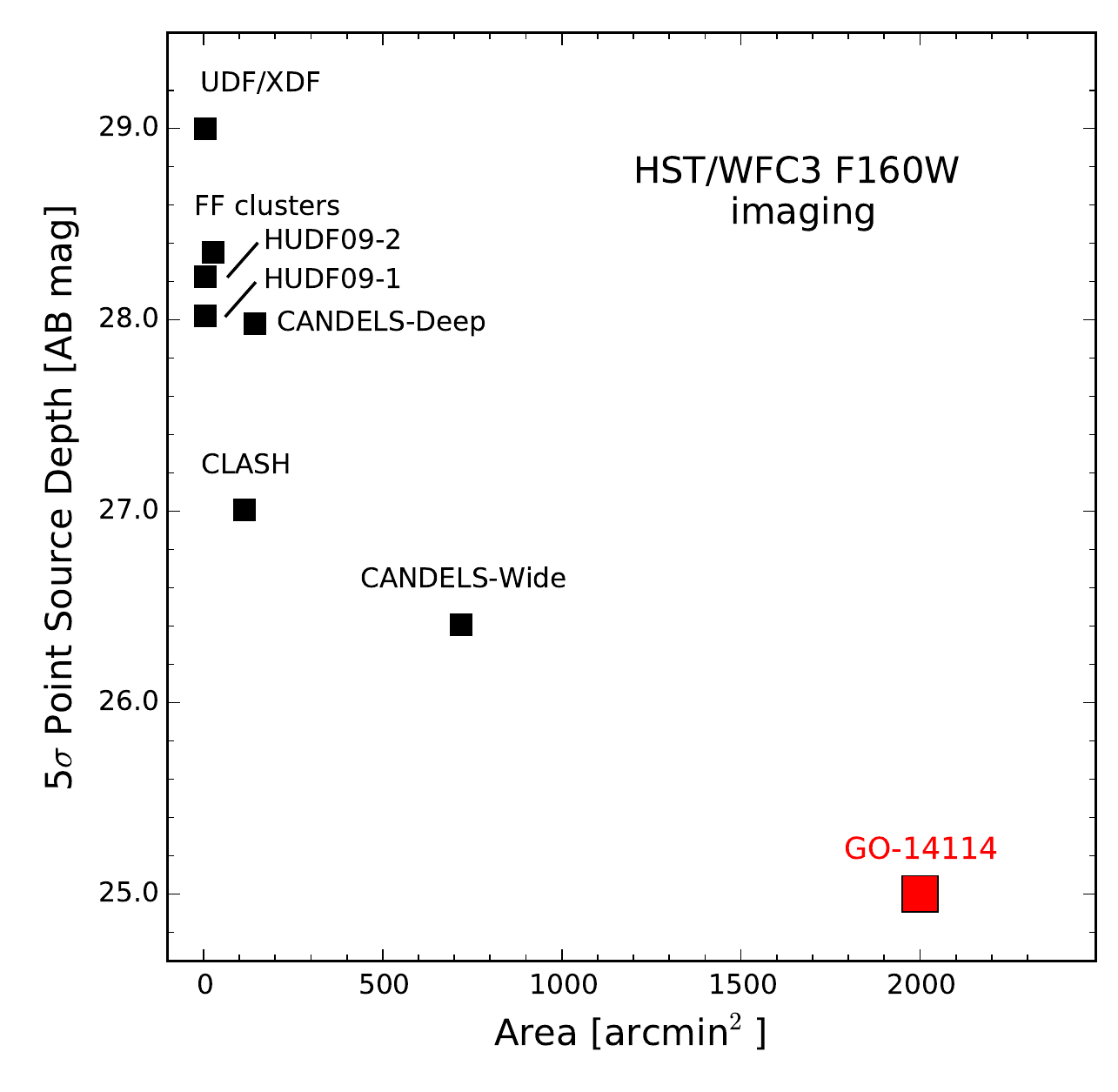}
\caption{\footnotesize Area vs. depth for WFC3/IR F160W imaging. The technique described in this paper as implemented in GO-14114 (red point) allows us to add a wide shallow tier to infrared surveys with HST. Furthermore,
the orbit total of
GO-14114 (57)
is an order of magnitude smaller than that of the other surveys
shown in this Figure.}
\label{fig:wedding_cake}
\end{figure}

Figure \ref{fig:wedding_cake} places \name in the context of other
WFC3 $H_{F160W}$ observations, in terms of their depth and covered
area. We probe a regime that was previously
unexplored with space-based IR imaging, despite the fact
that the orbit-total of GO-14114 is an order of magnitude smaller in terms of time allocation compared to, for example, COSMOS and CANDELS.

\section{Early Data}
\label{demo.sec}

We observed four orbits
of GO-14114 on UT 2015 October 15 and 16. The decision to obtain these
data was a trade-off: these early data allow us to assess the viability
of the methodology, but they come at a price.
The COSMOS field has a strongly variable near-IR background throughout
the year, as there is a relatively large amount of
zodaical dust in its direction. The background in October is so high
that the depth of the data is compromised; as we detail below these four
orbits are $\sim0.6$ magnitudes shallower than the rest of the
survey will be, when taken at lower background levels.
The four orbits (32
pointings) cover the top area of the middle ULTRAVISTA deep stripe,
North of the CANDELS field.

\subsection{Reduction and Analysis}
\label{sec:reduction}

We downloaded the raw and calibrated images from the Mikulski Archive
for Space Telescopes (MAST\footnote{\url{http://archive.stsci.edu}}).
The images were processed on the fly with
the best available calibration by the {\tt calwfc3} pipeline. In our
reductions we make use of both the flat-fielded final pipeline outputs
(\textsc{FLT}) and the calibrated intermediate MultiAccum exposures
(\textsc{IMA}). Each orbit consists of one guided and seven unguided
pointings. These two types of pointings are processed through similar
steps, but we note the differences where necessary.

We first carry out a basic reduction of the \textsc{FLT}, similar to
that described in \citet{skelton:14}.
We mask all sources in the
image and subtract a second order polynomial fit to the background. We
then align the image to an external reference world coordinate system
using the {\tt tweakreg} task. We use the $I_{F814W}$
images\footnote{\url{http://irsa.ipac.caltech.edu/data/COSMOS/images/acs_mosaic_2.0/tiles/}}
and catalogs provided by the COSMOS collaboration as a reference
\citep{koekemoer:07,massey:10}. As in \citet{skelton:14}, we use all
sources in the image for alignment,
not only stars.
Even though the \textsc{FLT} images are slightly smeared, we
prefer to use them for this rough alignment because the sources have higher signal to noise. A fine alignment step is performed
later in the reduction. The aligned \textsc{FLT}s are drizzled
together to create a preliminary mosaic for the orbit. We run {\tt
  SExtractor} \citep{Bertin96}
on this mosaic to create a segmentation map which is used
as a mask of sources in later steps of the reduction.

\begin{figure*}[htbp]
\centering
\includegraphics[width = 0.95\textwidth]{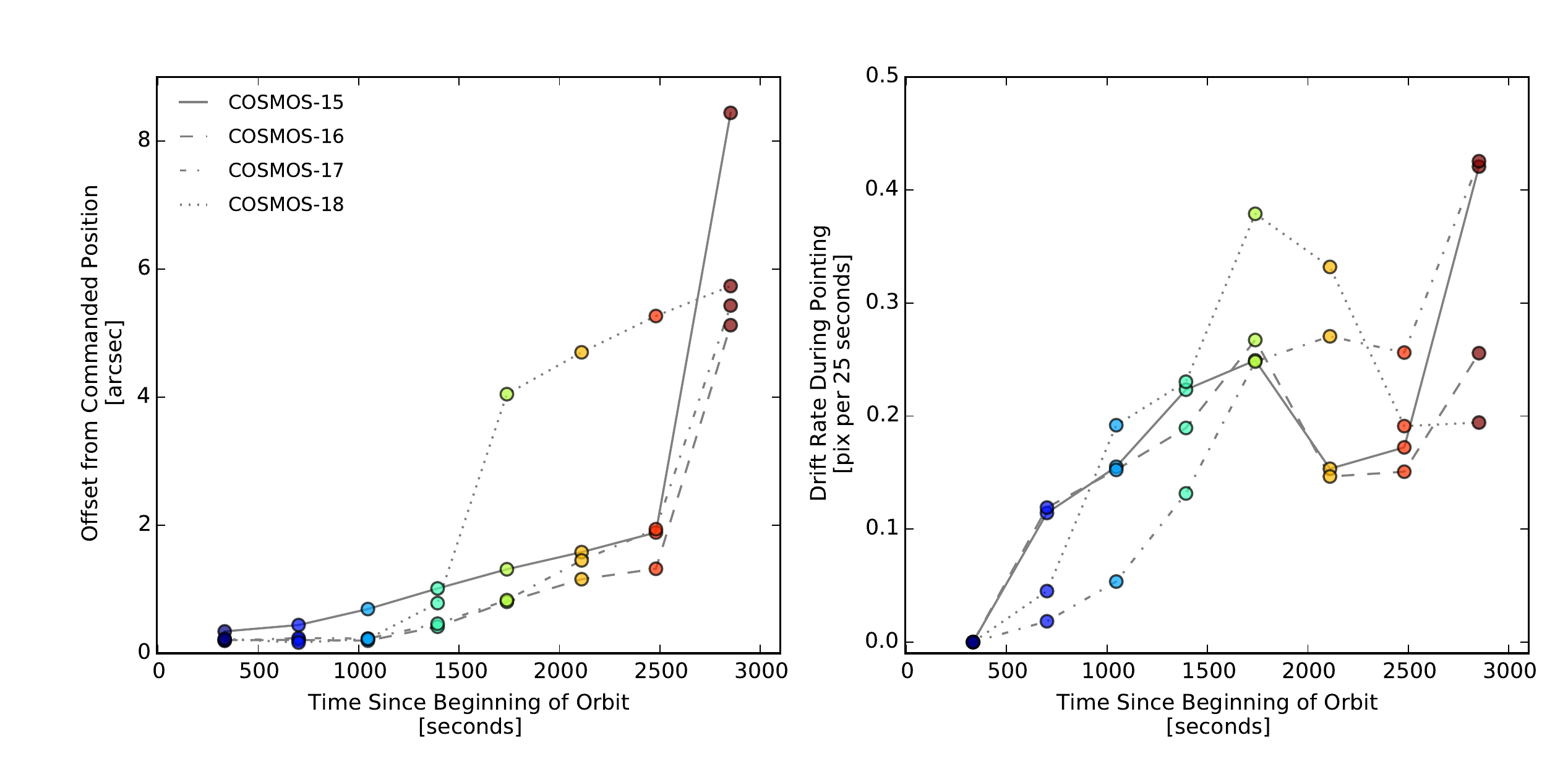}
\caption{\footnotesize {\it Left:} Offsets relative to the commanded
  position as a function of time for the eight exposures in each of
  the four observed orbits. The increase with time is expected,
  however the source of the large offsets at the end of the orbit (and
  halfway through COSMOS-18) is not yet understood. {\it Right:} Drift rate
  during the pointing in pixels per 25 seconds as a function of
  time. Following the first guided pointing (no drift), the drift rate
  increases with time.}
\label{fig:offsets_time}
\end{figure*}

The {\tt tweakreg} task provides the offsets of each of the 8
exposures from the commanded pointing position. 
Generally, the relative pointing accuracy of {\it HST} is very high, and
the RMS precision of offsets within an orbit is $\sim 2$ to 5 milli-arcsec with fine lock 
on two guide stars and when the offsets are small \citep{astrodrizzle}. 
In the absence of guide stars, we find that
the unguided exposures show
larger relative pointing errors, as large as $8\arcsec$. The left panel
of Figure \ref{fig:offsets_time} shows the offsets from the commanded
position as a function of time since the start of the orbit. The
offsets increase monotonically with time as can be expected from the
build-up of uncompensated forces on the spacecraft during the orbit.
Furthermore,
sudden jumps in pointing accuracy of $\sim4\arcsec$ can be seen
for the last position in COSMOS-15, 16 and 17. In orbit COSMOS-18 the
jump happens after the fourth position. The cause for these larger
offsets is not yet understood. The offsets are small relative to the instrument
FOV ($10\%$ of the WFC3 FOV); however users should make sure that there
is sufficient overlap between pointings to compensate for this effect
if a contiguous mosaic is needed.

Next, we proceed to reconstruct unsmeared images from the non-destructive
reads. These reads are preserved in the \textsc{IMA} images:
these are multi-extension files which contain all
the individual reads from the original exposure. Each read has been
bias-, overscan- and dark-current-subtracted as well as
flat-fielded. The up-the-ramp fits have also been carried out,
flagging cosmic rays. The \textsc{IMA} images are in units of
electrons per second. For each read $i$, starting with the second, we
subtract the preceding read, $i-1$. These difference images are treated
as the
science arrays in our analysis.
We construct an error array based on the read
noise, Poisson noise and flat field uncertainties. The data quality (DQ) array
is taken directly from the \textsc{IMA}.  These three arrays (science,
error, DQ) are paired with the header from the aligned \textsc{FLT}
image and saved as a new file. In this manner, each pointing is split
into 10 or 11 new 25-second images, depending on the number of
reads. Below we refer to the set of 10 or 11 such images produced from a
single original image as a ``sequence''. Guided and unguided exposures
are treated in the same manner.

We mask all sources (using the masks created from the \textsc{FLT}
files) and perform a median background subtraction for
each of these new images. As noted above,
the zodaical
background in the data is very high, at $\approx 2.2$\,e$^-$/s. Typical
values of the background in the CANDELS observations of the COSMOS
field range from 0.6\,e$^-$/s to 0.8\,e$^-$/s. For background levels below 
0.9\,e$^-$/s, a 250 second exposure would be read-noise limited (RN $\sim 15$ e$^-$), 
however, the current observations are background-limited.

The treatment of cosmic rays deserves careful attention. Cosmic
rays are flagged by the up-the-ramp fits of the {\tt calwfc3}
pipeline. However, there are two issues with this procedure. First,
due to the telescope drifts, objects move across the detector during
the exposure. This change in flux from one read to the next causes the
up-the-ramp fits to flag the varying pixels as cosmic rays. This issue
primarily affects point sources, where the size of the source is
commensurate with the drift during the exposure. To remedy this, we
reset all cosmic ray flags within the boundaries of objects (as
identified by the segmentation map).
This has the obvious negative effect of resetting the flags for
cosmic rays that fall on top of
objects. The second issue is that cosmic rays which
occur between the zeroth read (at 2.9 seconds) and the first read (at
27.9 seconds) cannot be identified by the ramp fits. While the first
issue affects only unguided pointings, the second affects all
pointings. In order to identify those cosmic rays, and also to flag
any cosmic rays which were erroneously reset in the previous step, we
use a combination of L.A.\,Cosmic \citep{lacosmic} (run on the FLTs) and
{\tt astrodrizzle} \citep[run when combining the individual reads;][]{astrodrizzle}. The 4096
flag is added to the DQ array for all pixels identified by both
methods. We visually verified that our adopted
procedure correctly identifies nearly all cosmic rays and leaves
the central pixels of compact objects intact.

\begin{figure}[htbp]
\centering
\includegraphics[width = 0.45\textwidth]{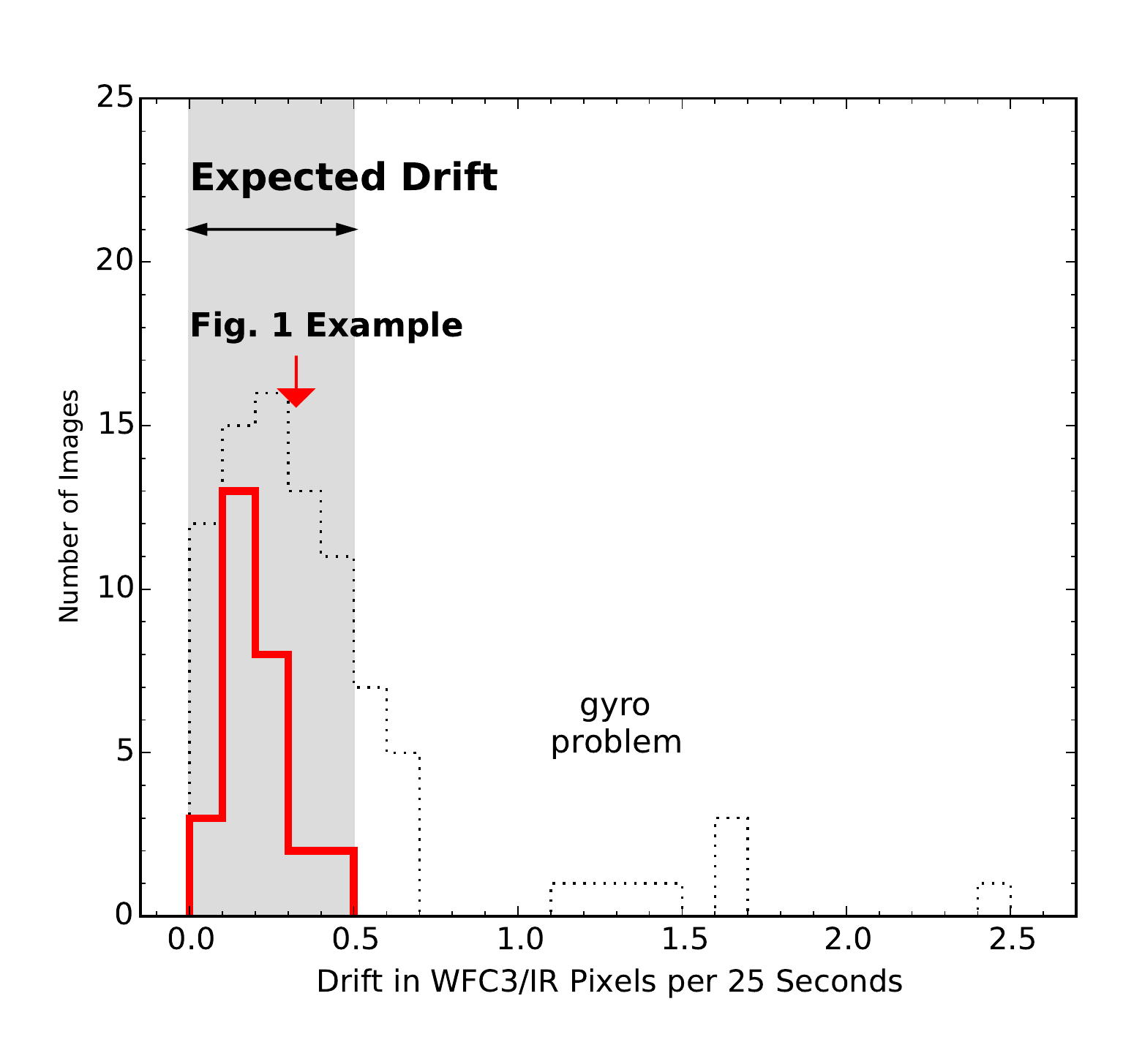}
\caption{\footnotesize Distribution of measured drift rates for the GO-14114 observations (solid red histogram) compared to those for 87
archival WFC3 exposures where
guiding failed (dotted black histogram). The drift rates are shown in units of pixels per 25 seconds, which measures the drift between reads in our program. The expected drift based on engineering predictions is up to $0\farcs 002$ per second (gray region) and our observations fall well within that. The pointing shown in Figure \ref{method.fig} is indicated with an arrow.}
\label{fig:drift_hist}
\end{figure}

\begin{figure*}[htbp]
\centering
\includegraphics[width = 1.1\textwidth]{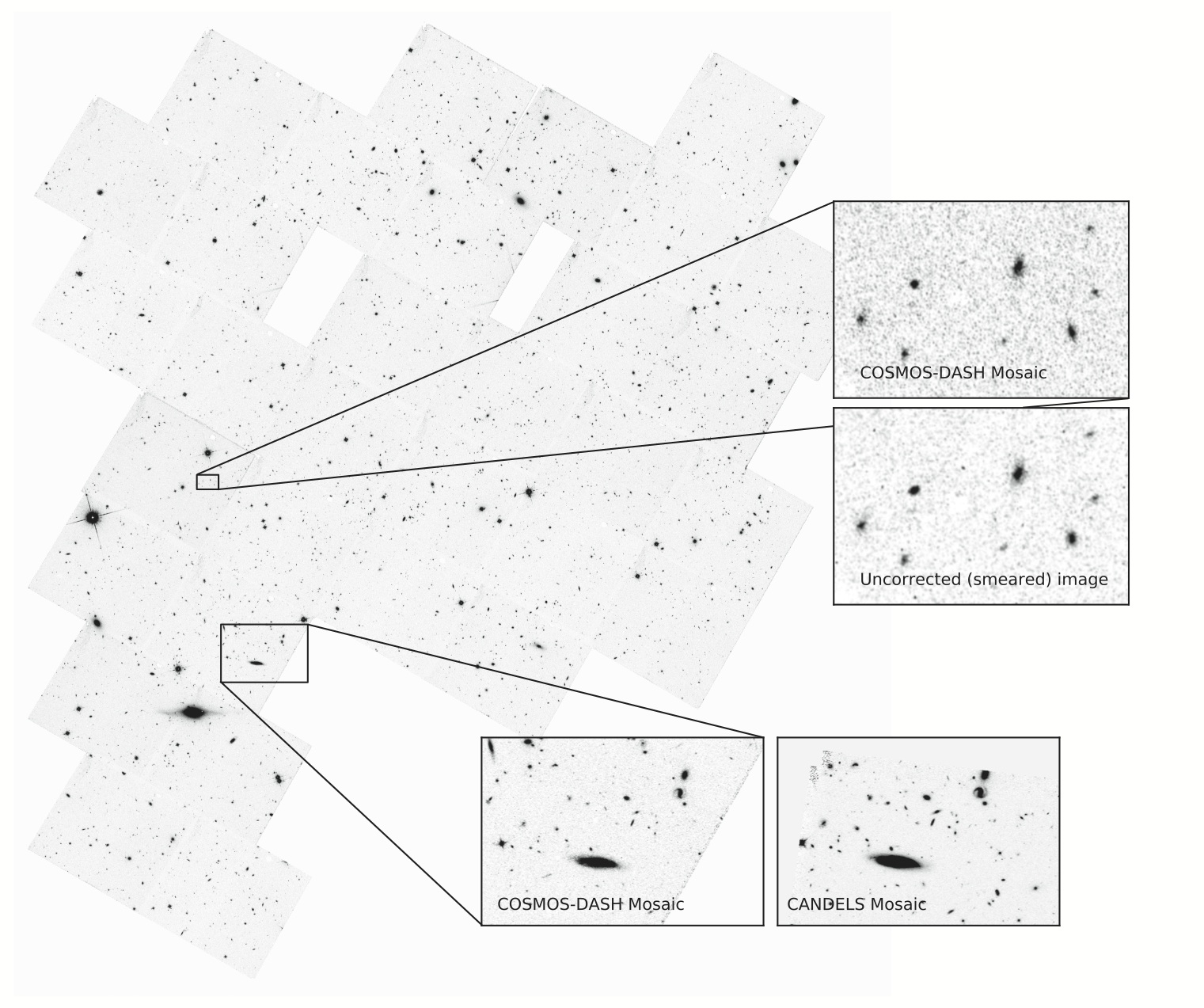}
\caption{\footnotesize The \name $H_{F160W}$ mosaic. The insets in the
top right corner show a portion of the mosaic in pointing 8
of COSMOS-15
before and after the alignment of the individual reads. The drift is
evident in compact objects but overall still very small. The
inset in the lower right part of the image shows a cutout of the
mosaic compared to the same portion of the CANDELS COSMOS mosaic
\citep{grogin:11,koekemoer:11,skelton:14}.}
\label{fig:mosaic}
\end{figure*}

\begin{figure*}[htbp]
\centering
\includegraphics[width = 0.95\textwidth]{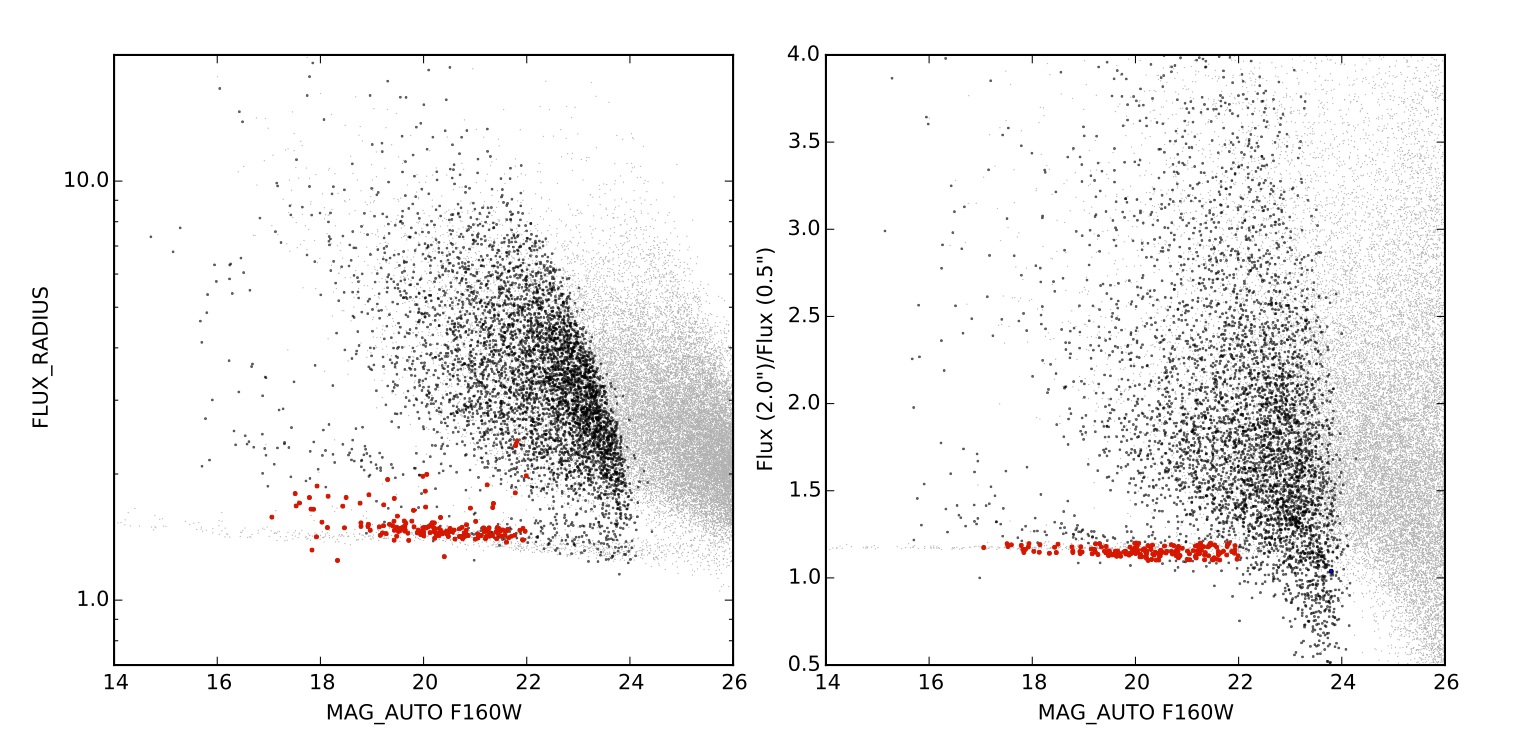}
\caption{\footnotesize SExtractor's FLUX\_RADIUS in pixels
  vs.\ MAG\_AUTO $H_{F160W}$ (black points) for the \name mosaic (left)
  and ratio in the fluxes measured in two different apertures
  (2\arcsec and $0\farcs 5$) vs.\ MAG\_AUTO $H_{F160W}$ (right). Stars
  that are used to construct the PSF are marked with red points (see text).
For comparison we also show the distribution of points in
  the CANDELS COSMOS mosaic of \citet[][gray points]{skelton:14},
  accounting for the different pixel scales. The stellar sequence in
  the \name mosaic is in the same location and as tight as the
  guided CANDELS COSMOS mosaic.}
\label{fig:mag_radius}
\end{figure*}

The images in all non-guided sequences are aligned to the COSMOS
$I_{F814W}$ mosaic. The alignment is performed with the {\tt tweakreg}
task in {\tt astrodrizzle} \citep{astrodrizzle}. Again, we use all sources within the
image to calculate the offsets. Here we only solve for $x$ and $y$ shifts
and do not fit for rotation or change in scale. The typical rms in the
differences of matched positions is $\sim0.4$ native pixels, identical
to that achieved when aligning the much deeper CANDELS COSMOS images
\citep{skelton:14}. For each sequence we calculate the offset between
the first and the last read to determine the mean rate of drift. In
the right panel of Figure \ref{fig:offsets_time} we show the drift
rate as a function of time for each of the four orbits. The drift rate
increases with time, as expected, but stays within
$0\farcs 002$ per second, or 0.42 pixels per 25 seconds. In
Figure \ref{fig:drift_hist} we show the distribution of drift rates
for all unguided sequences (solid red line) compared to archival WFC3 IR
observations that were taken in gyro-only
control (dotted black line). The mean drift rate in our
observations is $0\farcs 001$ per second or 0.2 pixels per 25 seconds,
smaller than that of the archival observations. It should
be noted that in the archival data the switch to gyro-control was not
deliberate, but the result of the failure to acquire (or the loss
of) 
a guide star. Our data suggest
that when the switch to gyro control is done deliberately, the drift
rates are very small and within the expected range.

Final mosaics are produced with {\tt astrodrizzle} using all sequence
images. We use exposure time weighting, a square kernel and
$pixfrac=0.8$.  We drizzle the images to a pixel scale of
0.1\arcsec.
The final mosaic of the current observations is $9100\times10200$ pixels, centered at
RA=$10:00:25.4$, DEC=$+2:34:51.2$. In Figure \ref{fig:mosaic} we show
the full mosaic. We also show a zoomed-in cut-out (top right) to
demonstrate the difference between the original drifted image and the
final processed image.  The example shown is from the eighth pointing
of COSMOS-15, i.e., one of the pointings with the largest drift rate
(Figure \ref{fig:offsets_time}). Another set of cut-outs (bottom
right) compare our mosaic to that from the CANDELS observations.
The mosaic is of high quality, with no obvious problems or defects:
it looks like a shallow version of the CANDELS imaging, as intended.
In the following subsections we characterize  the data quality.

%The data reduction and analysis tools described here are available for download through a Github repository\footnote{\url{https://github.com/ivastar/wfc3_gyro}}

%Initial processing, shifting, drizzling, mosaicing.

\subsection{Point Spread Function}

In order to assess the quality of the reconstructed data, we first
turn to the point spread function (PSF).
A key question is whether
the fact that the exposures were unguided has led to a net broadening
of the PSF compared to guided exposures. 

We construct
the PSF directly from
the $H_{F160W}$ mosaic using stars in the image.
We run {\tt SExtractor} on the final mosaic to detect objects and
perform photometry. Fluxes are measured in a series of 
apertures with different radii.
In the left panel of Figure \ref{fig:mag_radius} we show
the {\tt SExtractor} FLUX\_RADIUS (the radius of the circle centered
on the barycenter that encloses about half of the total flux)
vs.\ MAG\_AUTO (a total magnitude measurement). Point sources follow a
tight sequence with small sizes at all magnitudes and match well the
stellar sequence in the CANDELS COSMOS catalog of \citet[][gray
  points]{skelton:14}. Points that scatter above the stellar sequence
at bright magnitudes are saturated stars. In the right panel of Figure
\ref{fig:mag_radius} we show the ratio of the flux in a large aperture
($2\farcs 0$) to that in a small aperture ($0\farcs 5$) as a function
of magnitude which shows a similar tight sequence, which also matches
the CANDELS COSMOS catalog of \citet{skelton:14}.  The median FWHM of the 
stars in these sequences is $0\farcs21$ (2.1 pixels) compared to $0\farcs19$
 (3.2 pixels at $0\farcs06$/pix) from the CANDELS COSMOS mosaic \citep{skelton:14}, which is fully guided, indicating that we are recovering the original resolution. The sequences in
both panels of Figure \ref{fig:mag_radius} are useful diagnostics of
the image quality such that large spreads or offsets in the stellar
sequences can indicate problems with alignment or clipping of the
cores of stars. 

\begin{figure}[htbp]
\centering
\includegraphics[width = 0.45\textwidth]{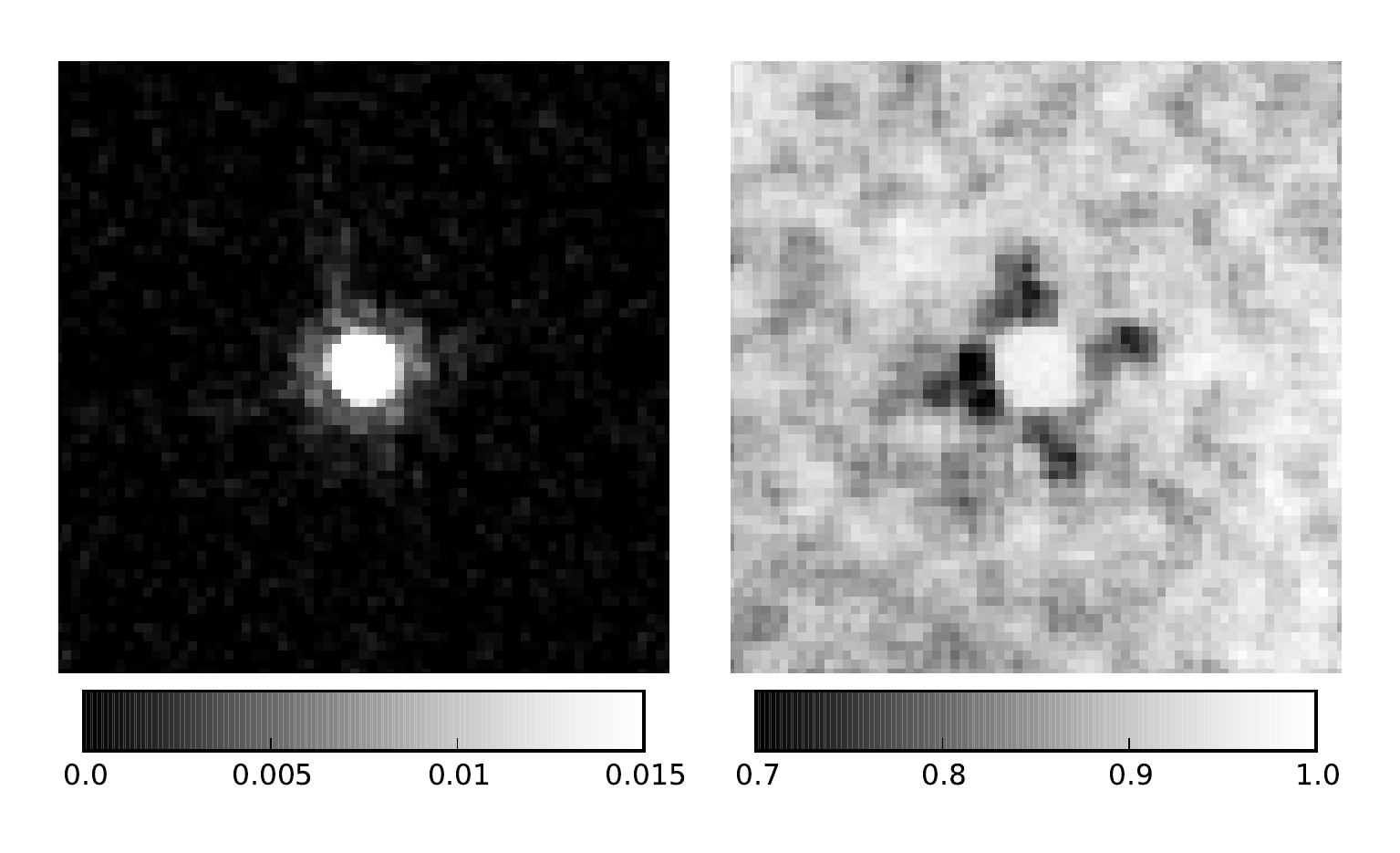}
\caption{\footnotesize Point spread function (PSF, left) constructed from stars within the \name mosaic and the corresponding combined weight map (right). Both the PSF and the weight map have been normalized by the peak value. The selection of stars is shown in Figure \ref{fig:mag_radius}. 106 stars with $H_{F160W}<22$ were used to create the PSF.}
\label{fig:psf}
\end{figure}

For a quantitative comparison, we follow the PSF construction method
of \citet{skelton:14}.  Stars are selected based the stellar sequence
in the right panel of Figure \ref{fig:mag_radius} such that the ratio
between the fluxes is $1.1<f(2\farcs 0)/f(0\farcs5) < 1.2$
(red points).
We visually inspect all selected stars and exclude 21 objects which
are either too close to the edge of the mosaic or have close
neighbors. Despite our effort, some stars still have suppressed weight at the center of their weight maps. 
We exclude 39 such stars, leaving a final sample of 106 objects with
$H_{F160W}<22$. Stars are distributed evenly across the field and,
since only 4 of the 32 pointings are guided, the final PSF will be
dominated by the unguided exposures. We mask neighboring objects
within a postage stamp cut-out of 84 pixels around each star.  The
postage stamps are recentered, normalized and then averaged to
determine the PSF and the PSF weight map (Figure \ref{fig:psf}).

\begin{figure}[hbtp]
\centering
\includegraphics[width = 0.45\textwidth]{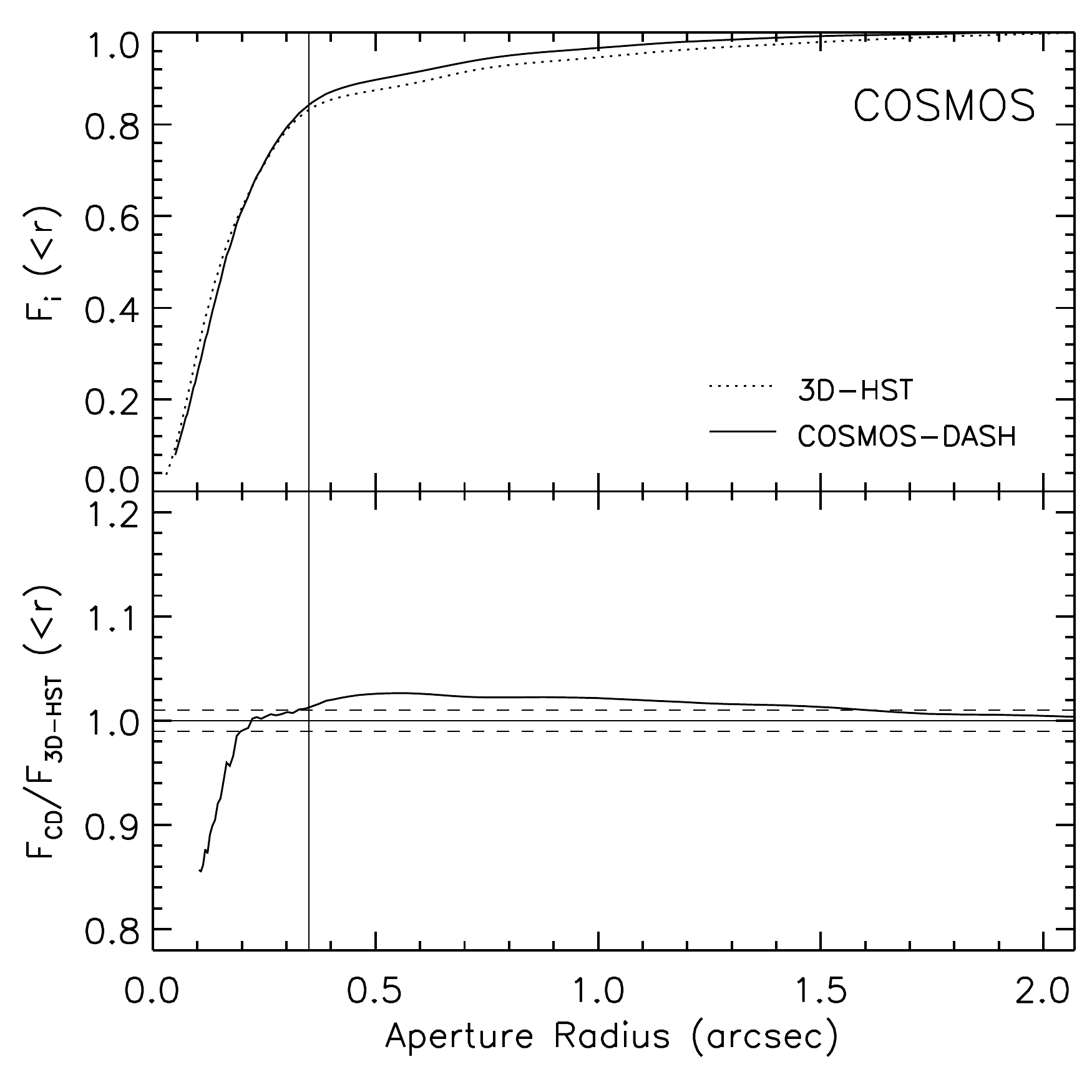}
\caption{\footnotesize Growth curve of the
point spread function.
{\it Top: } The fraction of light enclosed as a function of
radius relative to the total light within $r=2$\arcsec (solid
  curve). For comparison we also show the growth curve for the COSMOS
  CANDELS field constructed by \citet{skelton:14}. {\it Bottom: }
  Ratio between the \name and the \citet{skelton:14} growth curves as
  a function of radius. At the typical aperture used for photometry
  ($0\farcs 35$, vertical line) the differences between the two PSFs are $\sim1\%$.}
\label{fig:growth}
\end{figure}

The curve of growth, which is the fraction of enclosed light as a
function of aperture size (normalized at $2\farcs 0$), is shown in the
top panel of Figure \ref{fig:growth} (solid line). For comparison, we
also show the $H_{F160W}$ PSF derived from the CANDELS COSMOS
observations \citep{skelton:14}. The two growth curves show excellent overall
consistency.  A quantitative measure of the consistency is the ratio
between the two curves, which is shown
in the bottom panel of Figure \ref{fig:growth}. The \name
PSF has $\sim15\%$ less energy in the central pixel (for a pixel scale
of $0\farcs 1$ per pixel), however within the typical aperture used
for photometry, $0\farcs 35$ (solid vertical line), and out to
$2\farcs 0$ the differences between the two PSFs are $\lesssim1\%$. 
This demonstrates that by aligning the individual reads we have
recovered the resolution of guided {\it HST} images.

\subsection{Noise Characteristics}

Here we analyze the noise characteristics of the reconstructed images.
As a comparison sample we obtained from MAST all
other observations of extragalactic fields carried out with the SPARS25
sampling mode and matching the exposure times of our observations. In
total we find a total of 39 public datasets, containing 107 exposures
(\textsc{FLT}s) that match these criteria. We process these datasets
in a manner identical to the \name observations: we split them into
individual reads and drizzle those to a final distortion corrected
image (no alignment is necessary). In this analysis
we use the native pixel scale of $0\farcs 12$ per pixel.

\begin{figure*}[htbp]
\centering
\includegraphics[width = 0.75\textwidth]{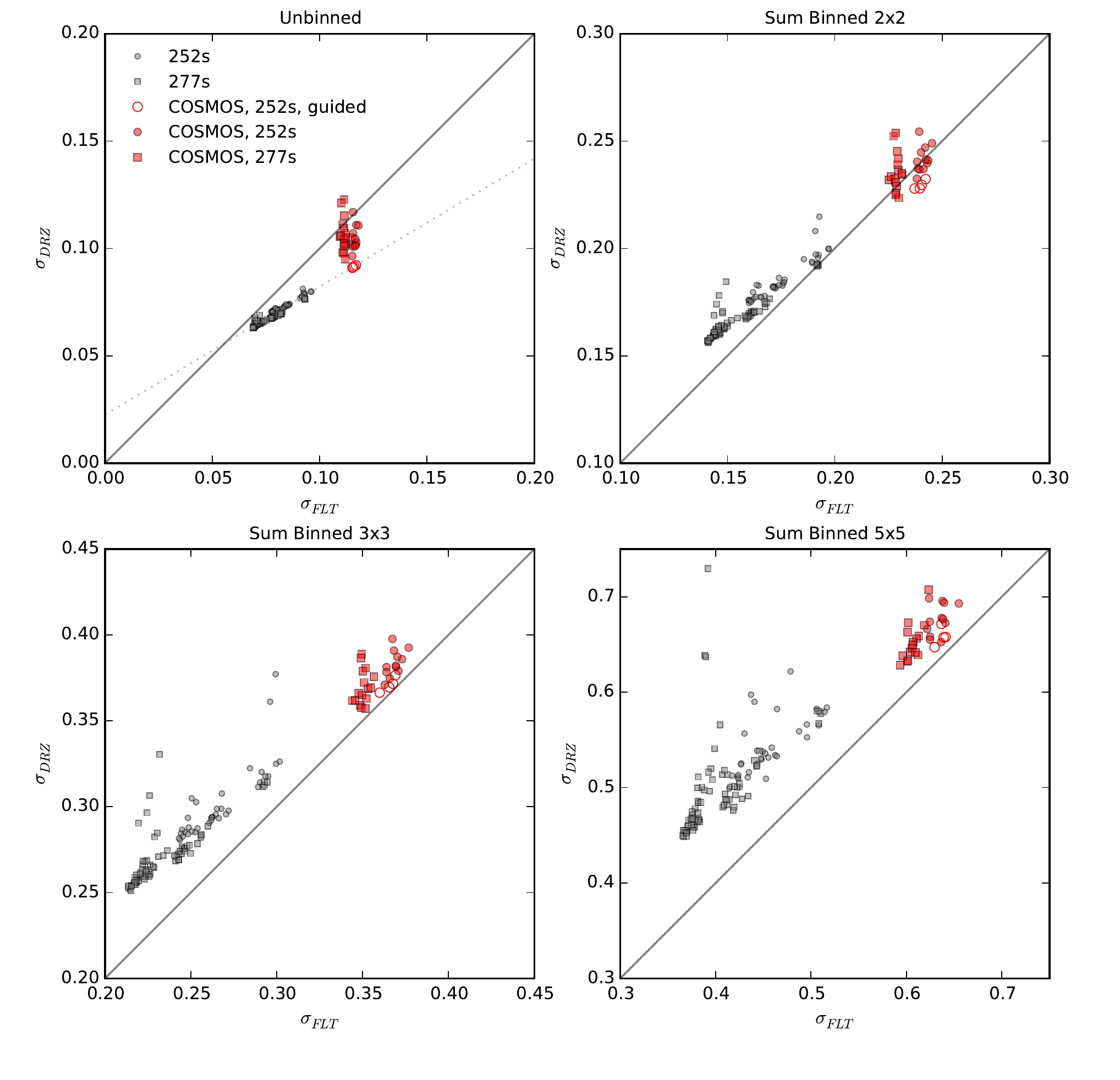}
\caption{\footnotesize Comparison between the pixel-to-pixel scatter in the noise of the \textsc{FLT} images, $\sigma_{\textsc{FLT}}$, and the noise of the drizzled individual reads, $\sigma_{\textsc{DRZ}}$. The left-most panel shows the noise at the native pixels of the image, while the remaining panels show how the noise changes when we rebin the images in coarser grids ($2\times2$, $3\times3$ and $5\times5$). The guided pointings (red open circles) have noise characteristics in agreement with analogous archival observations (gray symbols, dotted line fit), while the drizzled un-guided pointings have higher noise.}
\label{fig:sigma}
\end{figure*}

For each dataset we measure the pixel-to-pixel noise in the original
\textsc{FLT}s and in the final drizzled image, masking sources and
clipping the image edges. In the top left panel
of Figure \ref{fig:sigma} we
compare the noise in the \textsc{FLTs} to the drizzled
images. As expected the scatter is lower in the drizzled images:
drizzling artificially suppresses the noise in the images as a result
of the rebinning of flux \citep[][\S 3.3.1]{multidrizzle}. The \name
observations (red symbols) have higher noise overall as a result of
the elevated background levels (see \S \ref{sec:reduction}).
The four
{\em guided} pointings (red open symbols) follow the trend of the comparison
sample (broken line):
they behave in the same way as other guided data, except with
higher noise due to the increased background. The difference is
broadly consistent with the difference in the background level: the
zodaical background is a factor of $\sim 3$ higher than for typical
exposures, which translates into a factor of $\sqrt{3}$ higher noise.

However, the drizzled {\em unguided} pointings
(filled red symbols) exhibit higher
noise than the drizzled guided images. On average, the noise in the unguided
pointings is 15\% higher than in the unguided ones (and up to 28\%
higher). This corresponds to a loss in depth of 0.15 ( and up to 0.27)
magnitudes. The reason for this behavior is that, as a result of the
shifts within each pointing, a larger number of independent pixels
contribute to a single drizzled output pixel. 

This loss of depth due to the drift-and-shift method depends on the
spatial scale. It is rare that one is interested in individual
pixels; typical aperture photometry is performed over scales of
several pixels, to match the size of the PSF or to match the size
of spatially-extended objects. As an example, the aperture
used in the \citet{skelton:14} catalogs has a diameter of $0.7\arcsec$.
To assess the noise increase on larger scales we rebinned the
\textsc{FLT} and
\textsc{DRZ} images to coarser grids and re-measured the pixel-to-pixel
variation.
In the $2\times2$ binned images (Figure \ref{fig:sigma},
second panel), the noise in the unguided exposures is
on average 5\% higher than
that in the unguided exposures. In the $3\times3$ and
$5\times5$ binned images the noise is only $\sim3\%$ higher.
We conclude that, in ``real world'' applications, the loss in depth
due to our method is only $\sim0.03-0.07$
magnitudes.

Finally, we stress that {\em for these particular orbits} the depth
is limited by the high zodaical background, which results from our
decision to schedule them early in the Cycle.  A full analysis of the
depth of the \name data, including a completeness analysis,
will be performed when all data are taken in Fall 2016.

\subsection{Galaxy Structural Parameters}

A small section of the mosaic of 32
pointings overlaps with existing
$H_{160}$ data from the CANDELS survey. We measure structural parameters
of objects in this overlap region from our mosaic,
and compare them to measurements
of the same objects in CANDELS by
\citet{vdWel12}.
The same methodology is used here, and we refer to \citet{vdWel14}
for the details. Briefly, GALFIT \citep{galfit} is used to fit Sersic profiles \citep{sersic} to the images, using the GALAPAGOS wrapper
\citep{galapagos}. The local background is not a free parameter in the fit, but determined in
an initial step.

For the comparison we selected objects brighter than $H_{F160W} = 22$ and excluded objects with {\tt use\_phot = 0} which removes edge objects and stars  \citep{skelton:14}. We further excluded galaxies with uncertain GALFIT fits ($f\geq2$) in either catalog, corresponding to bad fits and no fits. The two catalogs are crossmatched based on position with a tolerance of 0\farcs5. No attempt is made to match the segmentation maps. The final sample consists of 48 objects.

\begin{figure*}[ht]
\centering
\includegraphics[width = 0.75\textwidth]{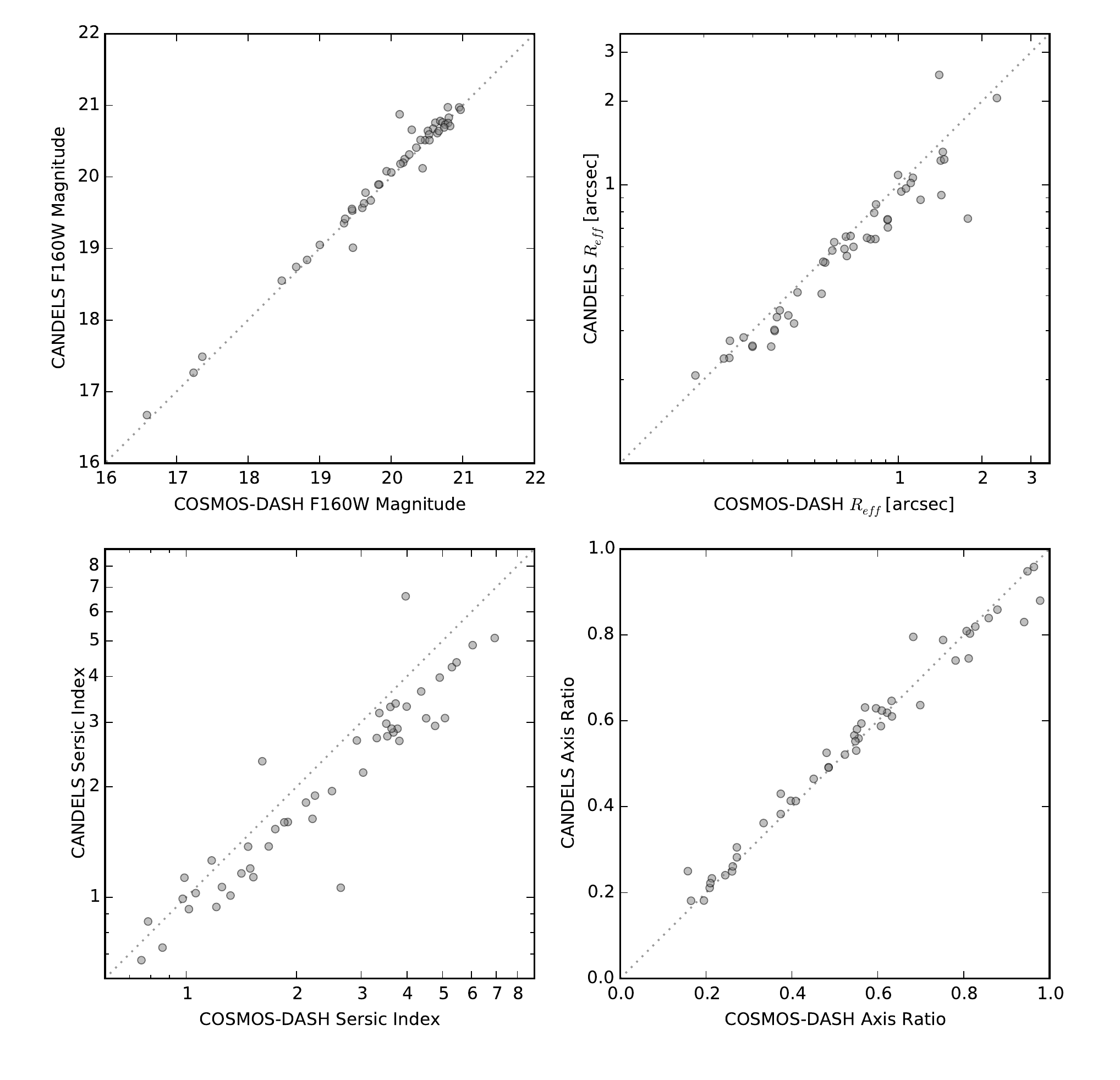}
\caption{\footnotesize Comparison between CANDELS and \name structural
parameters, as determined with GALFIT. Overall there is good agreement.
The sizes and Sersic indices show a small systematic effect, with the
sizes in \name\ larger than those in CANDELS by 10\% on average
(see text).}
\label{fig:struc}
\end{figure*}

Figure \ref{fig:struc} compares
the structural parameters measured in \name to those measured
in CANDELS. Overall there is excellent agreement. The total magnitudes
are offset by 0.05, with very small scatter. The median offset in
axis ratios is 0.004, again with almost no scatter ($\sigma \sim 0.02$).
There is a small systematic offset in sizes, which correlates with
a small systematic offset in Sersic indices. This offset is
negligible for galaxies with low Sersic indices and/or small sizes.
The median size
difference for all objects is 0.04 dex or 10\%, with objects in \name\ larger
than in CANDELS, and the scatter is 0.05 dex. It is not clear whether
this small difference is caused by a small remaining error in our
PSF, issues with the background subtraction, or other effects. There
is also the (perhaps unlikely) possibility that CANDELS slightly
underestimates the sizes.
Further analysis of these effects is limited by the high background
in these four orbits and the small number of stars that are suitable
for PSF reconstruction; we expect to explore this further when the
program is completed. Based on the data we have in hand, we conclude
that sizes can be estimated with an accuracy of $\sim 10$\,\%.

%\begin{figure*}[ht]
%\centering
%\includegraphics[width = 0.5\textwidth]{figures/mag_depth.pdf}
%\caption{\footnotesize Magnitude comparison between between the deep and shallow COSMOS observations.}
%\label{fig:mag_depth}
%\end{figure*}

\section{Conclusions}

In this paper we show that HST can obtain wide-field near-IR data
in a relatively small number of orbits. The ``drift and shift'' (DASH) technique
that makes this possible is well understood, and has now been demonstrated
to produce science-grade data at the full resolution of the WFC3 camera.
The low near-IR background from space means that such wide-field surveys
reach depths that are competitive with the deepest ground-based surveys,
despite the short $\sim 300$\,s per-pointing integration times.

Our medium-sized, 57-orbit Cycle 23 program covers an
area more typically associated with a Treasury survey. Were our
technique applied in an actual Treasury program, one could cover an
area of five square degrees in approximately 500 orbits. This opens up
regimes previously regarded as out of Hubble's reach, both for
``blank'' extra-galactic surveys and for
specific targets such as M31 and the Magellanic Clouds.

\begin{acknowledgements}
We thank Merle Reinhart for advice on the observation planning and Katherine Whitaker for help with the data reduction. This paper is based on observations made with the NASA/ESA Hubble Space Telescope, obtained at the Space Telescope Science Institute, which is operated by the Association of Universities for Research in Astronomy, Inc., under NASA contract NAS 5-26555. These observations are associated with program GO-14114. Support for GO-14114 is gratefully acknowledged.
\end{acknowledgements}

\bibliographystyle{apj}
\bibliography{cosmos_pasp}

\newcommand{\noop}[1]{}
\begin{thebibliography}{1}
\expandafter\ifx\csname natexlab\endcsname\relax\def\natexlab#1{#1}\fi

\bibitem[Abbott et al.(2005)]{des} Abbott, T., et al. 2005, 
arXiv:astro-ph/0510346

\bibitem[Arnaboldi et al.(2007)]{kids:07} Arnaboldi, M., 
Neeser, M.~J., Parker, L.~C., et al.\ 2007, The Messenger, 127, 28 

\bibitem[Barden et al.(2012)]{galapagos} Barden, M., 
H{\"a}u{\ss}ler, B., Peng, C.~Y., McIntosh, D.~H., 
\& Guo, Y.\ 2012, \mnras, 422, 449 

\bibitem[Bertin 
\& Arnouts(1996)]{Bertin96} Bertin, E., \& Arnouts, S.\ 1996, \aaps, 117, 393 

\bibitem[van Dokkum(2001)]{lacosmic} van Dokkum, P.~G.\ 2001, 
\pasp, 113, 1420 

\bibitem[Ellis et al.(2013)]{ellis:13} Ellis, R.~S., McLure, 
R.~J., Dunlop, J.~S., et al.\ 2013, \apjl, 763, L7 

\bibitem[Erben et al.(2013)]{cfhtlens} Erben, T., Hildebrandt, 
H., Miller, L., et al.\ 2013, \mnras, 433, 2545 

\bibitem[Fruchter, Sosey, et al.(2009)]{multidrizzle} Fruchter, A. and Sosey, M. et al.\ 2009,"The MultiDrizzle Handbook", version 3.0, (Baltimore, STScI)

\bibitem[Giavalisco et al.(2004)]{goods} Giavalisco, M., 
Ferguson, H.~C., Koekemoer, A.~M., et al.\ 2004, \apjl, 600, L93 

\bibitem[Gonzaga et al.(2012)]{astrodrizzle} Gonzaga, S., Hack W., Fruchter, A., and Mack, J. et al.\ 2012, "The Drizzlepac Handbook", version 1.0, (Baltimore, STScI)

\bibitem[Grogin et al.(2011)]{grogin:11} Grogin, N.~A., Kocevski, 
D.~D., Faber, S.~M., et al.\ 2011, \apjs, 197, 35 

\bibitem[Illingworth et al.(2013)]{illingworth:13} Illingworth, G.~D., 
Magee, D., Oesch, P.~A., et al.\ 2013, \apjs, 209, 6 

\bibitem[Jarvis et al.(2013)]{video} Jarvis, M.~J., Bonfield, 
D.~G., Bruce, V.~A., et al.\ 2013, \mnras, 428, 1281 

\bibitem[de Jong et al.(2013)]{kids:13} de Jong, J.~T.~A., 
Verdoes Kleijn, G.~A., Kuijken, K.~H., 
\& Valentijn, E.~A.\ 2013, Experimental Astronomy, 35, 25 

\bibitem[Koekemoer et al.(2007)]{koekemoer:07} Koekemoer, A.~M., 
Aussel, H., Calzetti, D., et al.\ 2007, \apjs, 172, 196 

\bibitem[Koekemoer et al.(2011)]{koekemoer:11} Koekemoer, A.~M., 
Faber, S.~M., Ferguson, H.~C., et al.\ 2011, \apjs, 197, 36 

\bibitem[Lotz et al.(2014)]{lotz:14} Lotz, J., Mountain, M., 
Grogin, N.~A., et al.\ 2014, American Astronomical Society Meeting 
Abstracts \#223, 223, \#254.01 

\bibitem[Massey et al.(2010)]{massey:10} Massey, R., Stoughton, 
C., Leauthaud, A., et al.\ 2010, \mnras, 401, 371 

\bibitem[McCracken et 
al.(2012)]{ultravista} McCracken, H.~J., Milvang-Jensen, B., Dunlop, 
J., et al.\ 2012, \aap, 544, A156 

\bibitem[Peng et al.(2002)]{galfit} Peng, C.~Y., Ho, L.~C., 
Impey, C.~D., \& Rix, H.-W.\ 2002, \aj, 124, 266


\bibitem[Rix et al.(2004)]{rix:04} Rix, H.-W., Barden, M., 
Beckwith, S.~V.~W., et al.\ 2004, \apjs, 152, 163 

\bibitem[{{Scoville} {et~al.}(2007){Scoville}, {Aussel}, {Brusa}, {Capak},
  {Carollo}, {Elvis}, {Giavalisco}, {Guzzo}, {Hasinger}, {Impey}, {Kneib},
  {LeFevre}, {Lilly}, {Mobasher}, {Renzini}, {Rich}, {Sanders}, {Schinnerer},
  {Schminovich}, {Shopbell}, {Taniguchi}, \& {Tyson}}]{scoville:07}
{Scoville}, N., {Aussel}, H., {Brusa}, M., {Capak}, P., {Carollo}, C.~M.,
  {Elvis}, M., {Giavalisco}, M., {Guzzo}, L., {Hasinger}, G., {Impey}, C.,
  {Kneib}, J.-P., {LeFevre}, O., {Lilly}, S.~J., {Mobasher}, B., {Renzini}, A.,
  {Rich}, R.~M., {Sanders}, D.~B., {Schinnerer}, E., {Schminovich}, D.,
  {Shopbell}, P., {Taniguchi}, Y., \& {Tyson}, N.~D. 2007, \apjs, 172, 1
  
  \bibitem[Sersic(1968)]{sersic} Sersic, J.~L.\ 1968, Cordoba, 
Argentina: Observatorio Astronomico, 1968,  

\bibitem[Skelton et al.(2014)]{skelton:14} Skelton, R.~E., 
Whitaker, K.~E., Momcheva, I.~G., et al.\ 2014, \apjs, 214, 24 

\bibitem[van der Wel et al.(2012)]{vdWel12} van der Wel, A., 
Bell, E.~F., H{\"a}ussler, B., et al.\ 2012, \apjs, 203, 24 

\bibitem[{{van der Wel} {et~al.}(2014){van der Wel}, {Franx}, {van Dokkum},
  {Skelton}, {Momcheva}, {Whitaker}, {Brammer}, {Bell}, {Rix}, {Wuyts},
  {Ferguson}, {Holden}, {Barro}, {Koekemoer}, {Chang}, {McGrath},
  {H{\"a}ussler}, {Dekel}, {Behroozi}, {Fumagalli}, {Leja}, {Lundgren},
  {Maseda}, {Nelson}, {Wake}, {Patel}, {Labb{\'e}}, {Faber}, {Grogin}, \&
  {Kocevski}}]{vdWel14}
{van der Wel}, A., {Franx}, M., {van Dokkum}, P.~G., {Skelton}, R.~E.,
  {Momcheva}, I.~G., et al., 2014, \apj, 788, 28

\bibitem[Williams et al.(1996)]{hdf} Williams, R.~E., 
Blacker, B., Dickinson, M., et al.\ 1996, \aj, 112, 1335 


\end{thebibliography}

\end{document}